\documentclass[amssymb,showpacs,pra,amsmath]{revtex4}

\usepackage{graphicx}
\usepackage{latexsym}
                   
\usepackage{latexsym}

\newcommand{\be}{\begin{equation}}
\newcommand{\ee}{\end{equation}}

\newcommand{\bex}{\begin{eqnarray}}
\newcommand{\eex}{\end{eqnarray}}
\newcommand{\bmin}{\begin{center}\begin{minipage}{460pt}}
\newcommand{\emin}{\end{minipage}\end{center}}
\newcommand{\bc}{\begin{center}}
\newcommand{\ec}{\end{center}}

\begin{document}

\title{The quantum measurement problem as a witness to ``It from bit''}
\author{R.    Srikanth}   \email{srik@rri.res.in}   \affiliation{Raman
Research Institute, Sadashiva Nagar, Bangalore.}


\begin{abstract}
A conceptual difficulty in the foundations of quantum mechanics is the
quantum  measurement  problem (QMP),  essentially  concerned with  the
apparent non-unitarity of the measurement process and the classicality
of macroscopic systems.  In an information theoretic approach proposed
by us  earlier \mbox{(Quantum  Information Processing 2,  153, 2003)},
which we  clarify and  elaborate here, QMP  is understood to  signal a
fundamental finite  resolution of quantum states,  or, equivalently, a
discreteness of Hilbert  space. This was motivated by  the notion that
physical reality is a manifestation of information stored and discrete
computations performed  at a  deeper, sub-physical layer.   This model
entails that states of sufficiently complex, entangled systems will be
unresolvable,   or,  {\em  computationally   unstable}.   Wavefunction
collapse  is  postulated  as  an  error preventive  response  to  such
computational  instability.  In  effect, sufficiently  complex systems
turn  classical  because  of   the  finiteness  of  the  computational
resources available to the physical universe.  We show that this model
forms a  reasonable complement to decoherence for  resolving QMP, both
in respect  of the problem of  definite outcomes and  of the preferred
basis  problem.  The  model  suggests that  QMP,  as a  window on  the
sub-physical universe, serves as a witness to Wheeler's koan ``it from
bit''. Some  implications for quantum computation  and quantum gravity
are examined.
\end{abstract}
\maketitle

\section{Introduction}
The quantum measurement problem (QMP)  is a key set of questions that,
arguably, every interpretation of  quantum mechanics must answer.  The
principal  problem  is  that  the wavefunction  in  quantum  mechanics
evolves  unitarily  according to  the  Schr\"odinger  equation into  a
linear superposition of different  states but measurements always find
the  physical  system  in  a  definite  state,  typically  a  position
eigenstate.  The  future evolution is  based on the system  having the
measured value at  that point in time, meaning  that measurement seems
to  affect the  system in  a way  not explained  by the  basic theory.
Formally,   measurement   precipitates   or   collapses   the   system
irreversibly,  probabilistically  and  non-unitarily into  a  definite
state.

Inspite  of the  fundamental nature  of QMP,  there appears  to  be no
unanimous agreement  as regards its resolution,  or even significance.
Various  interpretations or  models of  quantum measurement  have been
proposed to  resolve it.  The early  Copenhagen interpretation averred
that only  discussion about  probabilities was meaningful.   Rooted in
philosophical  positivism, this  view regarded  the wavefunction  as a
mathematical   tool  requiring   no  deeper explanation.    The
Copenhagen interpretation  appears to be more  of a way  to talk about
quantum mechanical  ``weirdness" in classical language,  rather than a
true interpretation  or model.  In  particular, its stark  contrast of
classical,  measuring  systems  from  the  measured,  quantum  systems
arguably only shifts QMP to a different base, without really resolving
it.

Now  it is  known  that  an inevitable  component  of the  measurement
process  is decoherence,  by which  a system  loses  coherence through
entanglement  with   the  measuring  apparatus   and  the  environment
\cite{max,zeh,kiefer,zur}.   Decoherence   is  sometimes  regarded  as
resolving  QMP,  though  it  is  not  clear  that  this  is  the  case
\cite{adl01}.  In a given  measurement whose outcome has been read-out
and is  thus known, the measured  system is formally  represented by a
pure  state.  Since  the  reduced density  operator  for an  entangled
system is necessarily mixed, a pure state cannot be entangled with any
other  system,   be  it  the  measuring  instrument   or  the  unknown
environment.   This  line  of  reasoning suggests  that  a  decohering
procedure  like  Eq.  (\ref{vonn})  below,  although  able to  explain
non-selective  measurements,  is   unable  to  account  for  selective
measurements,  that is, the  selection of  a single,  definite outcome
state \cite{sri03}.   To fully account for measurement,  it would seem
that decoherence  has to be complemented with  either a relative-state
interpretation  \cite{everett}   or  an  explicit   breakdown  of  the
superposition principle, which  is wavefunction collapse \cite{bas00}.
A problem with  the former interpretation is that it  does not seem to
lead to  the probabilistic nature  of the real-world  measurements.  A
difficulty with  collapse models is  that it may  look ad hoc,  if not
motivated on deeper grounds.

Detailed  models of  wavefunction collapse  that has  been extensively
studied.    In   the  dynamical   program   \cite{ghistan},  and   the
gravitational model by  Penrose \cite{pen96}, the projection postulate
is derived  as a  consequence of additional  physics.  In  the former,
this is achieved  via a small stochastic, nonlinear  term added to the
dynamical  equation of  the standard  theory;  in the  latter, via  an
energy uncertainty arising from gravitational field superpositions due
to  states that are  spatially apart.   The model  we present  here is
similar in  spirit, but  with state vector  reduction attributed  to a
certain  computational rather  than physical  or dynamical  cause.  It
aims to  resolve QMP within  the decoherence scenario. For  details of
various other approaches to resolve QMP, cf.  the introduction in Ref.
\cite{sri03}.

In Ref.  \cite{sri03},  we were led to the  position that wavefunction
collapse,  in conjunction  with decoherence,  is a  reasonable  way to
resolve QMP. In this work, we revisit and concretize some of the ideas
presented there.   To present QMP somewhat more  formally: suppose the
system  to  be  measured  is  in  the  state  $|\psi\rangle  =  \sum_j
c_j|j\rangle$.  Further suppose the complex of the measuring apparatus
and  the   environment  interacts   with  the  system   according  to:
$|j\rangle|R\rangle|E_0\rangle                          \longrightarrow
|j\rangle|m_j\rangle|E_j\rangle$,  where  $\{|R\rangle, |m_j\rangle\}$
is a  complete basis for the measuring  apparatus, and $\{|E_0\rangle,
|E_j\rangle\}$  a complete,  decohered basis  for the  environment.  A
puzzle  posed  by  QMP  is  that the  system's  measurement  does  not
discernably    lead    to    the    observation    of    superposition
$|\Psi^{\prime}\rangle$ obtained according to
\begin{equation}
\label{vonn}
|\Psi\rangle     \equiv     \sum_j    c_j|j\rangle|R\rangle|E_0\rangle
\longrightarrow        |\Psi^{\prime}\rangle       \equiv       \sum_j
c_j|j\rangle|m_j\rangle|E_j\rangle,
\end{equation}
as linearity suggests, but (as  far as the system-apparatus complex is
concerned)   probablisitically   to   {\em   one}  of   the   outcomes
$|j\rangle|m_j\rangle$.   Note  that,  non-selectively, the  procedure
(\ref{vonn})   leads   to   the   statistical   mixture   $\rho=\sum_j
|c_j|^2|j\rangle|m_j\rangle\langle j|\langle m_j|$.   To say that this
replacement of quantum  coherence with classical correlations explains
the classicality of the measurement outcomes misses the point of QMP!

Two problems are raised by QMP.  The first one concerns the absence of
coherent superpositions after measurement.  This is called the problem
of definite outcomes  (QMP1).  The second problem asks  how the choice
of particular basis  in which the `collapse' happens  is made. This is
the preferred basis problem (QMP2).

\section{The computational model for quantum measurement (CMQM) 
\label{sec:cmqm}}
The basic  philosophy behind our  present model, called  the computational
model for quantum  measurement (CMQM) \cite{sri03}, 
is that in  some sense physical
reality is fundamentally  informational and computational.  We picture
the  universe  as  composed  of  two  layers:  an  apparent,  physical
superstructure  (``physical  universe''),  and a  hidden,  algorithmic
matrix  (``sub-physical  universe'')  supporting  it. The  states  and
evolution  of systems in the physical universe
correspond  to information  stored and
computations  performed by the  sub-physical universe.   Physical laws
correspond to the underlying {\em discrete} sub-physical 
computational algorithms.

The idea that physical phenomena emerge from discrete informational or
computational  processes is  not  an entirely  new notion.   Wheeler's
phrase ``it  from bit"  first gave voice  to this  \cite{hweo}.  Refs.
\cite{cah}  and \cite{ulf02}  have  proposed ways  to obtain  physical
dynamics from discrete  algorithms. The novel feature of  our model is
that it connects the latter with QMP.  In this way, our model suggests
that  the  classicality  of  the  macroscopic world  is  a  window  on
discrete, sub-physical  information processing and  that therefore QMP
potentially  serves as  a  witness to  a  profound connection  between
physical reality and discrete, sub-physical processes.

We begin by  positing that quantum states in  finite dimension $D$ are
{\em algorithmically  bounded}, i.e.,  that the total  information (in
bits) required  to specify such a  state is finite.   We identify this
information with the  Kolmogorov complexity or algorithmic information
\cite{cha87}  of  the  state,  in  the  sense that  it  is  a  minimal
description of  the state with  respect to some fiducial  basis.  This
suggests that  there exists an  intrinsic limit $\mu$ to  the accuracy
with which a quantum state can  be specified.  We call $\mu$ the state
resolution  parameter.   An  immediate  consequence  is  that  quantum
operations are also  algorithmically bounded.  Algorithmic boundedness
is  based  on  the  constructivist  philosophy  that,  even  though  a
conventional  quantum state living  in a  Hilbert space  is in  fact a
platonic  entity requiring infinite  number of  bits to  be specified,
still any physical  instance thereof is always finite.   This limit is
understood to come from basic discreteness of Hilbert space itself, in
a  sense clarified  below.   A rather  mundane  interpretation of  the
algorithmic information of the quantum  state is that it is the amount
of memory  space at $\mu$-bit accuracy that  a sub-physical simulation
must allocate for a physical state.

More precisely,  we define the  algorithmic information for  states in
dimension $D$ to be the information:
\begin{equation}
\label{sv}
{\cal A}(D) \equiv - \sum p_j \log p_j = D\mu ~{\rm bits},
\end{equation}
where,  for simplicity,  we  have  ignored the  fact  that because  of
normalization, $(D-1)\mu$  suffice to  specify the state  at $\mu$-bit
accuracy.  Here logarithms  refer  to  base 2  by  default.  Thus  the
algorithmic  information for a  state depends  only on  its dimension,
with $\mu$ bits per amplitude  (dimension): $\mu/2$ for the real part,
and  $\mu/2$ for  the  imaginary  part.  In  contrast,  note that  the
maximum  accessible information, $\log  D$, is  exponentially smaller.
Algorithmic boundedness  can be given a  geometric interpretation.  If
we express the distance between states in a finite dimensional Hilbert
space  in terms  of  Hilbert space  angle,  a measure  of distance  on
projective Hilbert space, given by the Fubini-Study metric \cite{cav},
then the minimum resolvable  separation between two distinct states is
$2^{-\mu/2}$.   In  the  limit   $\mu  \to  \infty$,  we  recover  the
conventional continuum state space.

It is known  that the energy $E$ of the  system upper-bounds the speed
at which the system can perform {\em classical} computations, which is
roughly  $E/\hbar$ logical operations  per second  (ops) \cite{llo02}.
To  see this,  for instance  consider the  evolution of  a  qubit with
logical states $|0\rangle$ and $|1\rangle$ on which we perform the NOT
operation. To  flip the qubit one  can apply a  potential $\hat{H}_0 =
E_0|E_0\rangle\langle E_0|  + E_1|E_1\rangle\langle E_1|$  with energy
eigenstates  $|E_0\rangle =  (1/\sqrt{2})(|0\rangle +  |1\rangle)$ and
$|E_1\rangle   =   (1/\sqrt{2})(|0\rangle   -  |1\rangle)$.    Because
$|0\rangle = (1/\sqrt{2})(|E_0\rangle + |E_1\rangle)$ and $|1\rangle =
(1/\sqrt{2})(|E_0\rangle  - |E_1\rangle)$, each  logical state  has an
energy spread  $\Delta E  = (E_1 -  E_0)/2$. Under application  of the
potential,  the system prepared  in state  $|0\rangle$ after  time $t$
becomes:  $|\Psi(t)\rangle   =  \frac{1}{\sqrt{2}}\left(|E_0\rangle  +
e^{i2\Delta Et/\hbar}|E_1\rangle\right)$.   From this it  follows that
the  time  taken  to  flip  the qubit  to  $|1\rangle$,  and  likewise
vice-versa, is given by  $\pi\hbar/2\Delta E$.  One can similarly show
that AND  and FANOUT  gates, and  hence a set  of gates  universal for
classical   computation,  can   be  accomplished   in   about  similar
time. Therefore, the speed  of performing classical logical operations
is $f_{\rm classical}  \sim 2E/\hbar\pi$ ops, where we  set $E_0 = 0$.
Suppose  $n$ qubits with  total energy  $E$ process  some information.
Then each performs operations at rate $\sim 2E/n\hbar\pi$, so that the
total rate is $2E/\hbar\pi$, independent of $n$.

The implication  of algorithmic boundedness of operations  is that the
evolution  of  any  state  $|\psi\rangle$  is  equivalent  to  logical
operations performed by the sub-physical matrix at the finite rate:
\begin{equation}
{\cal    F}(D,E_j)     =    \frac{2^{\mu/2}}{\hbar}\sum_jE_j    \equiv
\frac{2^{\mu/2}D\bar{E}} {\hbar}~~{\rm ops}.
\label{eq:speed}
\end{equation}
This may be seen as follows. In the discretized state space, note that
the  $j$th amplitude  evolves in  time $2\pi\hbar/E_j$  through $2\pi$
radians.   That is,  in  the  complex plane  it  sweeps through  about
$2\pi\times 2^{\mu/2}$ cells, so that the rate per amplitude is $\nu_j
= 2^{\mu/2}E_j/\hbar$ ops, from  which we obtain Eq. (\ref{eq:speed}).
The   evolution  is  discrete.    Intuitively,  one   might  visualize
continuous evolution, where the state vector `snaps to' to the nearest
lattice  cell \cite{bun05}  in  discrete time-steps  executed at  rate
$\nu_j$.   Any  change  during  time  intervals  smaller  than  ${\cal
F}^{-1}$ is  deemed unresolved and undefined.   Thus, $\mu$ determines
both the precision  to which states, and their  unitary evolution, can
be resolved.  A rather mundane interpretation of the ${\cal F}(D,E_j)$
is that it  is the time-step rate in a  sub-physical simulation of the
evolution of state $|\psi\rangle$ at $\mu$-bit precision.

In  Eq.  (\ref{eq:speed}),  let  the $E_j$'s  be approximately  equal,
approximated by $E$. If the system  comprises of $N$ qubits, then $D =
2^N$, and the rate at which classical computations are performed is at
most about $NE/\hbar$.  On the  other hand, both ${\cal A}$ and ${\cal
F}$  scale  exponentially with  the  size  of  a composite  system  of
particles, in  contrast to classical information  storage capacity and
the maximum  classical computational rate.   It is simple to  see that
${\cal A}$  and ${\cal  F}$ by far  exceed the computational  power of
conventional  computers.  Suppose  we estimate  that there  are $10^9$
computers in all, each with a memory capacity of $10^{12}$ terabits, a
clockrate of $10^9$ Hz and  $10^5$ logical operations per clock cycle.
Therefore the  combined memory capacity and computational  rate of all
computers together is $10^{21}$  bits and $10^{23}$ ops, respectively.
For instance, if $\mu=50$ bits  (a better attempt to estimate $\mu$ is
discussed  later), from  Eq. (\ref{sv})  we see  that  these resources
suffice  to support  the state  of  no more  than 64  qubits, and  the
computational  resources for tracking  the evolution  of not  a single
qubit  more  energetic  than  a  cosmic  microwave  background  photon
(temperature $T \approx 2.8$ K).

With the  discretization of  state space ${\cal  H}$, it is  not clear
that  the mathematical  structure of  the discretized  state  space so
obtained,  denoted  ${\cal H}_{\mu}$,  is  strictly  a Hilbert  space,
because the  resolution-limited amplitudes do  not form a  field.  One
might consider  whether ${\cal  H}_{\mu}$ is a  vector space  over the
finite  field of Gaussian  integers \cite{gring}  modulo a  very large
prime $p$ of order $2^{\mu/2}$?  Probably not, because this can easily
be shown to lead to a situation where, given integers $a, b$ such that
$a <  b$, and $\psi \sim  a|0\rangle + b|1\rangle$, still  $a^2 > b^2$
(mod $p$), which is inconsistent with the probabilistic interpretation
of amplitudes.  There may be  no simple structure to ${\cal H}_{\mu}$.
Still,  as  a matter  of  terminology,  we  will usually  call  ${\cal
H}_{\mu}$ as discretized Hilbert space.  Also note that states are not
in general  truely normalized  for finite $\mu$.   Likewise operations
are not strictly unitary, but finite approximations thereof. A unitary
operation   $U$  in   ${\cal  H}$   is  replaced   by   its  $\mu$-bit
discretization, the `$\mu$-unitary' operation $U_{\mu}$.

One might be concerned about  the consistency of such an approximation
scheme.   One worry  might  be  that $SU(N)$  group  structure of  the
rotation of  an $N$-level system may  not be obtained as  the limit of
ever larger  finite discrete subgroups.  However, there  are models in
which continuous symmertry (rotational or Lorentz) is recovered in the
long  wavelength limit  from  underlying dynamics  with only  discrete
symmetry,  an  immediate   example  being  lattice  QCD  \cite{bun05}.
Another example is that of a model of spinless point particles hopping
on a flux lattice, which  gives rise to low-energy excitations obeying
the Dirac equation \cite{zee}.

Similar results  concerning discreteness can  be deduced also  for the
Schr\"odinger equation.   To see this,  we only note that  there exist
finite  simulation algorithms for  classical digital  computers, which
compute discrete  valued $\psi(x)$ in  discrete space and  time steps,
and  can approximate  continuum Schr\"odinger  evolution  to arbitrary
accuracy.   We therefore  regard the  $\mu$-limited  quantum mechanics
(QM$_{\mu}$)  formally  as  a  sub-physical  simulation  of  continuum
quantum  mechanics  (QM),  consuming  finite computational  power  and
memory, and  parametrized by finite constant $\mu$.   One then regards
conventional  ${\cal  H}$ as  the  continuum  approximation to  ${\cal
H}_{\mu}$.

\subsection{An entanglement monotone as measure of system complexity}
Given  a  collection of  objects,  the  total algorithmic  information
${\cal A}$ required to describe  it depends on whether the objects are
entangled  or  not. In  particular,  ${\cal  A}$  will depend  on  the
combinatorics of  the entanglement between the various  objects in the
collection.  For example, the  algorithmic information to describe two
separable  objects of  dimensions  $D_1$ and  $D_2$ is  $(D_1+D_2)\mu$
bits. If  now the two objects  interact to become  entangled, then the
combined  system's  updated  algorithmic  information  is  $D_1D_2\mu$
bits.  More generally,  if $N$  initially separable  $D$-level systems
become  entangled, correspondingly  ${\cal A}$  rises from  $DN\mu$ to
$D^N\mu$.  The  question  of   detecting  entanglement  in  a  general
multi-partite state  is still an  open question. Fortunately,  we need
concern ourselves only with the simpler, pure state entanglement. This
is  because the  sub-physical matrix  always `knows'  the state  it is
simulating, as it were.

It may  turn out that the  entangled state does  not resolvably differ
from a separable state, and is thus effectively separable at $\mu$-bit
precision. We need  a basis-independent way of describing  how a state
may  resolvably  differ from  a  separable  state.   Note that,  given
$|\psi\rangle = \sum_{j=0}^{D^N-1}\alpha_j|j\rangle$ that lives in the
space  of  pure  states  of  $N$  $D$-dimensional  objects,  which  is
$\mathbb{C}^D\otimes\cdots\otimes\mathbb{C}^D$,  not all of  the $D^N$
complex  parameters  have   nonlocal  significance.   Two  states  are
equivalent modulo local operations as far as their nonlocal properties
are concerned if they may be  reached from each other by local unitary
transformations, given by the  group $U(D)^N$, or $U(1)\times SU(D)^N$
if only independent effects  are considered. Each equivalence class of
nonlocally equivalent  states is an  orbit of this group.   Hence, the
space                   of                  orbits                  is
$\frac{\mathbb{C}^D\otimes\cdots\otimes\mathbb{C}^D}{U(1)\times
SU(D)^N}$  \cite{linp}.   From  this   we  find  that  the  number  of
independent  nonlocal (real)  invariants $\tau(\alpha_j)$  under local
unitary rotations must be
\begin{equation}
D^{N+1}-(D^2-1)N-1.
\label{eq:noofinv}
\end{equation}
A state would be deemed resolvably entangled if the $\tau(\alpha_j)$'s
differ sufficiently from their  separable values. This does not appear
to be a simple prescription for entanglement resolvability in terms of
amplitude   resolvability,   as   the  $\tau(\alpha_j)$'s   can   have
complicated functional forms. A simpler method is suggested below.

Consider      the     $N$-particle      state      $|\Psi\rangle     =
\sum_{\alpha_1,\cdots,\alpha_N}
c_{\alpha_1,\cdots,\alpha_N}|\alpha_1,\cdots,\alpha_N\rangle$.      Let
the   set  of  all   particles  be   $T  =   \{1,2,3,\cdots,N\}$.  Any
non-vanishing proper  subset of  $T$ is denoted  $y$. That is,  $y \in
{\cal T} \equiv 2^T - \emptyset  - T$. Denote $\overline{y} \equiv T -
y$.  Single particle marginal density matrices are denoted $\rho_j$.

A simple entanglement monotone for an $N$-partite pure state is:
\begin{equation}
\xi^{(N)} = \left\{
\begin{array}{ll}
\sum_{j=1}^N S(\rho_j)  & {\rm ~~~~if~} S(\rho_y) \ne  0 ~~\forall~~ y
\in {\cal T}. \\ 0 & {\rm ~~~~otherwise},
\end{array}\right. 
\label{eq:pan}
\end{equation}
where $S(\rho_y)  = -{\rm Tr}[\rho_y\log(\rho_y)]$ is  the von Neumann
entropy     and     $\rho_y     \equiv     {\rm     Tr}_{\overline{y}}
(|\Psi\rangle\langle\Psi|)$.    That    $\xi^{(N)}$   is   indeed   an
entanglement monotone  follows from  the fact that  marginal entropies
$S(\rho_j)$  do  not increase  under  local  operations and  classical
communication. The above definition is based on, but differs from, the
entanglement  measure   given  in  Ref.   \cite{pan04}   in  that  Eq.
(\ref{eq:pan})  does  not  reduce   to  entropy  of  entanglement  for
bipartite  states,   but  to  twice  that  value.    In  general,  Eq.
(\ref{eq:pan}) yields  $N\log D$,  and not $\log  D$, for  a maximally
entangled $N$-partite state.  This is convenient for our present need.

The advantage of  definition Eq. (\ref{eq:pan}) is that  it suggests a
direct extension to resolvable entanglement  for QM$_{\mu}$.  It is
well known that any bipartite  system can be Schmidt decomposed into a
state summed over a single  index \cite{nc00}. Precisely if the system
is  separable, its  Schmidt number  (number  of terms  in the  Schmidt
decomposition, which is  not larger than the dimension  of the smaller
of its two constituent sub-systems) is 1. The Schmidt coefficients can
be  obtained  directly  as  the  eigenvalues of  the  reduced  density
operator of  either sub-system. Let $(\lambda_+)_y$  denote the second
largest eigenvalue  of $\rho_y$ (or $\rho_{\overline{y}}$).   If for a
particular   bi-partition   $(y,    \overline{y})$,   we   find   that
$(\lambda_+)_y <  2^{-\mu/2}$, then the entanglement  between the sets
$y$ and $\overline{y}$ is deemed  unresolvable. The two parts are then
effectively  separable.  On  the  other hand,  if  $(\lambda_+)_y  \ge
2^{-\mu/2}$, then the entanglement is deemed resolvable, and the parts
$y$ and  $\overline{y}$ are  said to be  resolvably entangled  to each
other.

Thus,   we   define   $\mu$-bit   resolvable   (or   $\mu$-resolvable)
entanglement by:
\begin{equation}
\label{entpan0}
\xi^{(N)}_{\mu} = \left\{
\begin{array}{ll}
\sum_{j=1}^N  S(\rho_j)  &  {\rm  ~if~} (\lambda_+)_y  \ge  2^{-\mu/2}
~\forall~ y \in {\cal T}. \\ 0 & {\rm ~~~~otherwise}.
\end{array}\right. 
\end{equation}
Two systems  that are  not $\mu$-resolvably entangled  are said  to be
$\mu$-separable,  that  is,  separable  at  $\mu$-bit  accuracy.   The
discretized evolution  $U_{\mu}$ corresponding to  a unitary operation
$U$  is   resolvably  entangling  if,  acting  on   systems  that  are
effectively separable, it can produce $\mu$-resolvable entanglement.

\subsection{Finite quantum parallelism}\label{subsec:finpar}
A consequence of finite $\mu$ is that, if we assume that QM$_{\mu}$ is
consistent  and a  reasonable approximation  of QM,  then there  is an
upper bound, $D_{\rm max}$, to the dimension of the Hilbert space of a
{\em monolithic} system.  Such a  system is defined as either a single
fundamental object (whatever it may  be), or an entangled composite of
two or more such objects.  That is, by definition, a monolithic object
is not composed of two or more $\mu$-separable objects (fundamental or
otherwise). Suppose  an isolated, monolithic system exists  in a state
$|\psi\rangle   =  \sum_{j=0}^{D-1}   a_j|j\rangle$,  in   some  basis
$\{|j\rangle\}$.   If dimension $D  > 2^{\mu}$,  then it  follows that
there  exists at  least  one $j$  in  this basis  such  that $|a_j|  <
2^{-\mu/2}$, and therefore cannot be  {\em resolved}, even to its most
significant digit. This holds true for any other basis. Therefore, the
{\em  effective} dimension  $D_{\rm eff}$  of an  isolated, monolithic
system in QM$_{\mu}$ must satisfy:
\begin{equation}
D_{\rm eff} \le 2^{\mu},
\label{eq:deff}
\end{equation}
or,  equivalently, ${\cal  A}  \le 2^{\mu}\mu$  bits.  Any  coherently
evolving state  in ${\cal H}_{\mu}$ is  therefore {\em algorithmically
bounded}.

Physically,  this means that  the coherent  evolution of  any physical
system can  proceed along  at most $2^{\mu}$  parallel superpositional
pathways   (terms  in  a   coherent  superposition).   Thus,  infinite
parallelism  in a  continuum ${\cal  H}$  is replaced  by {\em  finite
parallelism} in ${\cal H}_{\mu}$.  Let us consider a monolithic system
accessing $D  < 2^{\mu}$,  which, by an  abrupt absorption  of energy,
would  have required  access to  $D^{\prime} >  2^{\mu}$ in  QM.  CMQM
posits  that  in  such  a  situation, the  non-resolvability  of  some
amplitudes leads to loss of  information from the system's state.  The
subsequent  evolution of the  object's state  depends on  whether this
lost information is  {\em significant} or not, in  the sense clarified
below.

Further,  consider  a  system  with  finite  average  energy  $\langle
E\rangle = \sum_j p_jE_j$,  where $p_j \ge 2^{-\mu}$ and $\left|\sum_j
p_j -  1\right| \sim O(2^{-\mu})$.  It  follows that all  $E_j$'s ($ 1
\le j  \le 2^{\mu}$) are finite.  Therefore,  along each computational
pathway,  any  such  system  evolves   at  a  finite  speed  given  by
$2^{\mu/2}E_j/\hbar$  ops.  As  a  result, finite  parallelism in  the
discrete  Hilbert space  entails that  any coherently  evolving system
corresponds to a finite rate of logical operations along finitely many
computational pathways in the  sub-physical matrix. In this sense, the
quantum universe is not {\em  computationally dense}, both in time and
in  ${\cal H}$.   Formally,  the sub-physical  matrix `simulates'  the
physical  evolution  (of  systems)  of  the  physical  universe  in  a
truncated basis of dimension at most $2^{\mu}$ and at finite speed. In
view of Eq.   (\ref{sv}), to any operation ${\cal  U}_{\mu}$ in ${\cal
H}_{\mu}$ is  associated an algorithmic information  of $D^2\mu$ bits,
i.e.,  the  dimension  of $U$  in  ${\cal  H}$  times $\mu$.   In  the
Heisenberg picture, we have that ${\cal U}_{\mu}$ is updated at a rate
of about $2^{\mu/2}D^2\overline{E}/\hbar$ ops, where $D \le 2^{\mu}$.

Even for modest values of $\mu$, such as say 100 bits, low dimensional
systems can hardly be  distinguished from the continuum case. Further,
if the Hilbert space dimension of the universe (= $\exp(S/k_B)$, where
$S$ is entropy  of the universe and $k_B$,  Boltzmann's constant) were
much less  than $2^{\mu}$, then the  effect of finiteness  of $\mu$ is
not  expected to  show up  easily at  any scale.   Yet,  clearly, even
familiar   systems   are   conventionally   considered   as   infinite
dimensional,  e.g.,  a  coherent  state of  light,  $|\alpha\rangle  =
e^{-|\alpha|^2/2}\sum_{n=0}^{\infty}(\alpha^n/\sqrt{n!})    |n\rangle$.
CMQM implies that the $n^{\rm  th}_{r}$ and later terms in the summand
will be unresolved, where $n_{r}$  $(\le 2^{\mu})$ is the smallest $n$
such        that         $2^{-\mu/2}        >        e^{-|\alpha|^2/2}
(|\alpha^{n}|/\sqrt{n_{r}!})$.   For example,  a  coherent pulse  with
$\alpha=2$ gives $n_r  = 45$, so that the  effective dimension $D_{\rm
eff}  \le 45$.  By  the remainder  theorem for  the Taylor  series for
$f(x)=e^{-\alpha^2}\exp(x=\alpha^2)$,  we   find  that  the  truncated
series sums  up to not  more than $e^{-65}$.   On the other  hand, the
average  probability  for  the  included  terms  is  about  $1/45  \gg
e^{-65}$.    Since   the   total   `loss   of   probability'   through
non-resolution (the truncation error) is much smaller than the average
probability in  each included superposition  term, the unresolvability
of the amplitudes for $n >  n_r$ is deemed {\em insignificant} and can
be ignored.   It is understood  that what is  usually taken to  be the
state   $|\alpha\rangle$   is   in   fact   physically   realized   as
$|\alpha_{\mu}\rangle    \equiv    \sum_{n=0}^{\le    n_r}    (\lfloor
e^{-|\alpha|^2/2}\alpha^n/\sqrt{n!}\rfloor) |n\rangle$, where $\lfloor
y\rfloor$ is  the $\mu$-bit rounded amplitude of  the system. Clearly,
it is difficult to practically distinguish $|\alpha_{\mu}\rangle$ from
$|\alpha\rangle$.   However, there  are situations  where the  loss of
amplitude information can be significant, and can thus not be ignored,
as discussed below.

\subsection{Computational instability}\label{subsec:comstab}
A set  of objects  may each satisfy  Eq. (\ref{eq:deff}), and  yet, by
interacting  via interactions that  are algorithmically  bounded, they
may   still  give   rise  to   a  monolithic   system   that  violates
Eq. (\ref{eq:deff}) and can  thus result in unresolvable amplitudes. A
simple example  is of  $N$ effectively $D$-dimensional  particles such
that $D \ll 2^{\mu}$, but $D^N  > 2^{\mu}$. Of course, the latter fact
by  itself  does  not  imply  that  the  there  is  {\em  significant}
unresolvability of amplitudes.

If  these $N$  particles are  separable, then  the  unresolvability is
statistically insignificant  because each monolithic  (separable) unit
within  the  system satisfies  Eq.  (\ref{eq:deff}).  But  if all  the
particles  become strongly entangled,  from Eq.  (\ref{eq:noofinv}) it
follows  that  all  the  $\sim  D^{N+1}$  nonlocal  invariants  differ
significantly  from  their  separable  values  and  hence  nearly  all
amplitudes are significant. The  loss of probability that would result
through non-resolution will thus be substantial. Therefore, the strong
interaction  regime, in  which $\xi^{(N)}_{\mu}  \longrightarrow N\log
D$, entails significant unresolvability.

We can describe  the passage from the resolvable  to the significantly
unresolvable situation using parameter  $\chi$, defined as the largest
entanglement $\xi^{(j)}_{\mu}$ ($j  \le N$) for any subset  of the $N$
particles in question:
\begin{equation}
\chi  \equiv \max_y \xi^{(|y|)}_{\mu}~~\forall~~y  \in ({\cal  T} \cup
T).
\label{eq:chi}
\end{equation}
The  condition  for  the  unresolvability  to  be  statistically  {\em
insignificant} is therefore:
\begin{equation}
\chi \ll \mu.
\label{eq:res}
\end{equation}
In the fully separable regime, where each of the $N$ particles forms a
`separable  island'  in a  pure  state,  $\chi  = 0$,  satisfying  Eq.
(\ref{eq:res}). But in the strong interaction regime, which results in
near-maximal entanglement and  nearly all amplitudes are statistically
significant,   $\chi    \longrightarrow   N\log   D    >   \mu$,   and
Eq. (\ref{eq:res}) fails.  At  this critical point, the system becomes
{\em  computationally unstable},  in the  sense that  the sub-physical
simulation of  the system becomes potentially very  lossy.  A physical
system  is  computationally  stable  only  if  Eq.  (\ref{eq:res})  is
satisfied.

A primary  element of CMQM is  that `collapse of  the wavefunction' or
`reduction of  the state vector'  is the error preventive  response of
the  sub-physical matrix  to computational  instability.  Wavefunction
collapse is modelled as  a highly discontinous transition during which
$\chi$ abruptly shifts  from about $\mu$ to 0 or  a value much smaller
than  $\mu$,  as the  system  is projected  from  a  state of  immense
entanglement  to  a  product  state  (though the  latter  may  not  be
separable in terms of the most basic degrees of freedom).

This  is postulated to  occur throught  the following  two-step random
procedure.   As  a  system  of  $N$  objects  becomes  computationally
unstable, any one of the objects, which we call the trigger, collapses
by  a random  projection into  a  basis whose  selection is  clarified
below.   Its  state  vector  thus  products  out  from  the  remaining
objects'.   Simultaneously, the  latter are  projected into  the state
correlated  with that  of  object $k$.   The  full collapse  therefore
consists   of  the  initial   trigger-collapse,  and   the  subsequent
correlated collapse.  The choice of the post-collapse state is assumed
to be random subject to the Born probability rule. For the case of Eq.
(\ref{vonn}),      the      final      state      can      be      any
$|j\rangle|m_j\rangle|E_j\rangle$  with  probability  $|c_j|^2$.   The
present model does not explain the origin of this randomness, which is
taken to be a fundamental feature.

There are several features of the model that are novel to the issue of
QMP. First is the feature that wavefunction collapse is related to the
finiteness  of  memory and  computational  capacity  available to  the
universe. Physically, this corresponds  to the discreteness of Hilbert
space.   Our   model  suggests   that  wavefunction  collapse   is  an
algorithmic rather than dynamical phenomenon. By the term algorithmic,
as against dynamical, is  meant that wavefunction collapse corresponds
rather to  discrete computations and a re-setting  of memory registers
in the sub-physical matrix,  than to a conventional Hamiltonian-driven
evolution.  We  venture to suggest that  computational instability and
wavefunction  collapse  in CMQM  are  analogous  to {\em  segmentation
fault} \cite{segfu}  and {\em crash} \cite{crash}  of ordinary, digital
computer  programs.    As  in   a  crash,  wavefunction   collapse  is
characterized by loss of information, corresponding to the destruction
of  nonlocal correlations.  

A  further  possibility  is  that  wavefunction  can  be  interpretted
physically as an  abstract phase transition, with $\chi$  as the order
parameter.  The phase transition is characterized by symmetry breaking
as  the state  vector jumps  from a  space of  larger symmetry  of the
highly  entangled, computationally unstable  state to  a one  of lower
symmetry.

\subsection{Algorithmic minimization}
The above analysis did not  address the preferred basis problem, QMP2.
The latter's  resolution in CMQM  relies on the  fact that there  is a
unique basis that minimizes the average algorithmic information of the
post-collapse  state. For  the  $N$-particle computationally  unstable
system, let  the $k$th particle  be the trigger object  projected into
some basis  $\beta = \{\beta_j\}$.  Let the  corresponding ensemble of
states of  the remaining objects obtained by  correlated projection be
$\phi=\{\phi_j\}$.  As  a simple example,  consider the (unnormalized)
maximally entangled state of three qubits, $|\psi\rangle = |000\rangle
+ |111\rangle$,  with $k=0$.  For  $\beta=\{|0\rangle,|1\rangle\}$, we
find  $\phi =  \{|00\rangle, |11\rangle\}$,  with $\overline{\chi}=0$;
for  $\beta=\{|\pm\rangle\}$,   we  find  $\phi   =  \{|00\rangle  \pm
|11\rangle\}$,  where $|\pm\rangle  = |0\rangle  \pm  |1\rangle$, with
$\overline{\chi}=2$.  If we require  that the average $\chi$ of $\phi$
should be minimized for the  post-collapse ensemble, it is easily seen
that for the state $|\psi\rangle$, the first basis is preferred.

More generally, suppose a  computationally unstable system is given by
the    state   $|\psi\rangle   =    \sum_j   c_j|j_0\rangle|j_1\rangle
\cdots|j_{N-1}\rangle$.   Without  loss  of  generality,  setting  the
trigger coordinate  $k=0$, suppose that  this particle collapses  in a
basis   $\{|m\rangle\}$  other   than   $\{|j_0\rangle\}$,  given   by
$|j_0\rangle  = \sum_{m}\alpha_{jm}|m\rangle$.   A  projection of  the
trigger  into an eigenstate  in the  $\{|m\rangle\}$ basis  leaves the
remaining particles in the entangled state (apart from a normalization
factor) $\sum_j c_j\alpha_{jm} |j_1\rangle\cdots|j_{N-1}\rangle$ and a
corresponding  $\overline{\chi} \approx  (N-1)\log D$.   On  the other
hand,   if  the   trigger  basis   is  $\{|j_0\rangle\}$,   we  obtain
$\overline{\chi}=0$. It is  obvious that this holds true  for any $k$.
We thus see  that the bases of objects in  which their entangled state
can  be  expanded  through   a  single  index  minimizes  the  average
algorithmic information of the  resulting ensemble.  Formally, this is
equivalent    to   measurement   in    the   basis    $\{\hat{P}_j   =
(|j_0\rangle\langle  j_0|)\otimes  (|j_1\rangle\langle  j_1|)  \otimes
\cdots \otimes  (|j_{N-1}\rangle\langle j_{N-1}|)\}$, which  is unique
to the state  $|\psi\rangle$ as the basis that  permits a single index
expansion of the  latter.  No matter what the  trigger coordinate, the
result is a projection of $|\psi\rangle$ in the basis $\{\hat{P}_j\}$.

Generalized to any system, this forms the {\em algorithmic minization}
principle  of  CMQM.   Formally, consider  ensemble  $\phi=\{\phi_j\}$
correlated   with  the  projection   of  the   trigger  object   in  a
computationally unstable  system, in basis $\beta  = \{\beta_j\}$.  We
denote the  average resulting entanglement  by $\overline{\chi}(\beta)
\equiv \sum_j\chi(\phi_j)$.   The {\em algorithmically  minimal basis}
$\gamma =  \{|\gamma_j\rangle\}$ is defined as the  one that minimizes
$\overline{\chi}$:
\begin{equation}
\overline{\chi}(\gamma) = \min_{\beta}\overline{\chi}(\beta).
\label{eq:minchi}
\end{equation}
For  states of  the form  (\ref{vonn}),  which are  quite general  for
measurements in the von Neumann measurement paradigm \cite{von32}, the
basis  $\{|j\rangle,  |m_j\rangle,  |E_j\rangle\}$ uniquely  satisfies
Eq. (\ref{eq:minchi}).

Thus, apart from computational instability and probabilistic collapse,
the third  main element  of CMQM is  that the final  (possibly random)
state following collapse is  chosen from the (algorithmically) minimal
basis.  This  is a `reasonable'  response to the  information overflow
experienced  during computational  instability.   Crucially, it  helps
resolve  the  preferred basis  problem  (QMP2)  as  it singles  out  a
specific basis in which the system is actualized or macro-objectified.
The particular element in the minimal basis which the system collapses
to is randomly chosen, subject  to the condition that probability $p_j
=   {\rm  Tr}(\hat{P}_j|\psi\rangle\langle\psi|)$.   If   we  restrict
attention  to a  subsystem  in a  monolithic  system, the  subsystem's
evolution is given  by a completely positive map  on density operators
\cite{nc00}.

As  the particles  in  the system  remain  dynamically interacting,  a
collapse is followed by  an episode of $\mu$-unitary evolution, during
which interaction  re-entangles the system,  making it computationally
unstable  again. This  is succeeded  by a  collapse, and  so  on.  The
perpetual cycle of  alternating collapses and $\mu$-unitary evolutions
gives rise  to a classical behaviour.   To see this, note  that in the
continuous  limit,  assuming  Markovian (time-local)  conditions,  the
collapse  of  any (open) sub-system  can  be  represented  by the  action  of
Lindblad  operators.  This  results  in an  evolution  of the  density
operator    described    by    a   Lindblad-type    master    equation
\cite{kiefer,zeh}, which  often suffices  to explain the  emergence of
macro-scale  classical   behaviour.  In  particular,   this  can  also
elucidate why position often emerges as the preferred basis.

The connection of our model to decoherence is worth stressing.  Notice
that the form Eq.  (\ref{vonn}), in which the states $\{|E_j\rangle\}$
are   practically  orthogonal,   results  from   decoherence.    As  a
consequence,     the     system-apparatus     complex    loses     all
coherence. Non-selectively, the density  operator of this subsystem is
the  same as  would  be  obtained by  projective  measurements in  the
$\{|E_j\rangle\}$   basis.     Therefore,   statistically,   CMQM   is
indistinguishable  from  decoherence.   Again,  in  the  more  general
context of evolution  of macroscopic open systems, we  noted that CMQM
yields  a Lindbladian  evolution.  Here  too, the  effect of  the CMQM
scenario is identical  to that due to decoherence.   However, CMQM has
the added feature of being  able to explain the appearance of specific
outcomes,  and can  thus serve  as  a complement  to decoherence  that
terminates the measurement process.

CMQM implies  that the parameter $\mu$ determines  the Heisenberg cut,
the mesoscopic  threshold presumably separating  the quantum microcosm
from the classical macrocosm. If $\mu$ were larger, then computational
instability  would be  attained  later than  earlier,  and so  quantum
superpositions would be seen at larger scales.  According to CMQM, the
classicality of the macro-world is due to the `accident' that $\mu$ is
too  small  in   comparison  with  the  degrees  of   freedom  of  the
universe. Following Ref. \cite{llo02}, suppose that the entropy $S$ of
the  universe is  $10^{120}$. The  corresponding dimension  is $D_{\rm
univ}  =  \exp(S/k_B)$. According  to  CMQM,  if  $2^{\mu} \gg  D_{\rm
univ}$, then even  the macroscopic world would be  quantum rather than
classical.   Conversely, the  fact that  the macro-world  is classical
therefore implies that $\mu < 10^{143}$.

In fact, $\mu$  is probably much smaller. We  suggest that experiments
of the  type studied in Refs. \cite{lucia,bouw}  (and references cited
therein) can possibly help determine the value of $\mu$ by identifying
the  mesoscopic  scale  at  which  quantum  behaviour  transitions  to
classical. However,  it should be noted that  these experiments cannot
be directly  used for  our purpose. They  rely on  identifying quantum
behaviour interferometrically  and thus do not  distinguish between an
actual collapse and the mere  loss of quantum coherence, i.e., between
selective  and  non-selective measurements.   As  a  result, they  are
insensitive to  the difference between  the effect of  decoherence and
that of  decoherence terminated by  a CMQM collapse.  We  believe that
modifications of such experiments can nevertheless be used to quantify
$\mu$.

\section{Links to other fundamental problems}\label{sec:lynx}
We note two consequences of CMQM.  First, there is an asymptotic limit
to  the power of  quantum computation.  Consider, for  example, Shor's
algorithm for  prime factorization \cite{shor}. To  factorize a number
$n$, we  choose a number $a$ that  is co-prime to $n$  and produce the
entanglement    $n^{-1/2}\sum_j   |j\rangle|0\rangle   \longrightarrow
n^{-1/2}\sum_j  |j\rangle|f(j)\rangle$,  by  way  of  determining  the
period  of the function  $f(j) =  a^j \mod  n$. However,  according to
Eq. (\ref{eq:res}), we should have  $n \le 2^{\mu}$, which is thus the
largest number that can be factorized using this algorithm.  A quantum
computer  that attempts  to  access higher  dimensions will  collapse,
losing  coherence.  Note that  according to  CMQM classicality  of the
macro-world itself is due  to interactions leading physical objects to
attempt  to  access larger  than  $2^{\mu}$  dimensions. The  physical
universe  can  thus  be  regarded   as  a  quantum  computer  that  is
dimensionally    too    rich,    and    turning   classical    as    a
consequence.  Interestingly, a different  approach to  discreteness of
Hilbert space limiting  the power of quantum computers  is reported in
Ref. \cite{bun05}.

Second,  we remark on  some connections  to quantum  gravity.  General
Relativity  (GR)  is well-tested  at  macroscopic  scales.  Yet,  when
extrapolated to very smaller  scales, it encounters inconsistencies in
the  form  of  singularities.   For  this, among  other  reasons,  one
suspects that  GR is not  universally valid, but that  at sufficiently
small scales, a theory of quantum gravity would be required.  CMQM can
be  motivated  along similar  lines,  by  arguing  that the  universal
validity  of quantum  mechanics at  all amplitude  scales  would imply
macroscopic  superpositions, contrary to  observations, and  that this
calls for new physics at very small (but significant) amplitudes.  The
loop  quantum gravity  \cite{smo04} approach  predicts  that spacetime
does not form a continuum, but is discrete.  This guarantees avoidance
of classical  singularities as well as the  high frequency divergences
of quantum field theory.  Similarly, CMQM requires the discreteness of
the  space  of  states,  which  guarantees  avoidance  of  macroscopic
superpositions by precluding arbitrarily massive quantum parallelism.

To manifest  the possible  granularity of space  we require  very high
energies in order to  probe Planck length phenomena.  Analogously CMQM
implies  that to  manifest  the granularity  of  state space,  massive
superposition  (entanglement)  is needed.   Yet,  a dramatic  contrast
between  these two  kinds of  granularity is  that whereas  the former
requires exotic  conditions (Planck scale energies)  to be manifested,
the latter is almost ubiqutous and inescapable: in the classicality of
the world we ordinarily see around us.

We claim that finite $\mu$  implies discreteness of spacetime, an idea
quite familiar in certain  approaches to quantum gravity, notably loop
quantum gravity, as  noted above.  At a given time,  let us consider a
cubic region of space, $R$,  of length $L$, that is sufficiently small
that the wavefunction  $\psi$ of a particle hardly  varies over it. To
begin with  consider space  as divied into  finite number of  cells of
size $\Delta  x$.  The  number of degrees  of freedom  in $R$ is  $N =
(L/\Delta x)^3$. If  the total probability of finding  the particle in
$R$ is $p  \le 1$, then the average amplitude  $\alpha$ in this region
satisfies  $\alpha \le \sqrt{p/N}  = \sqrt{p}(\Delta  x/L)^{3/2}$.  As
$\Delta x \longrightarrow 0$,  we have $\alpha \longrightarrow 0$.  In
particular,  if $N  > 2^{\mu}$,  then  $\alpha <  2^{-\mu/2}$ and  the
amplitudes in  the lattice in  the region $R$ are  unresolvable.  This
argument can  be applied  to every other  similarly chosen  region $R$
where the  particle has some  resolvable probability to be  found.  We
therefore require that $\Delta  x > L/2^{\mu/3}$.  The discreteness in
spacetime need not imply regularity.  Space need not be a lattice, but
might  be given by  a probabilistic  distribution consistent  with the
demands of Relativity theory.  The  latter condition will also imply a
corresponding discreteness in time, which by the way is also suggested
by  Eq.   (\ref{eq:speed}).   Conversely,  one  can  also  motivate  a
discreteness  of Hilbert  space, starting  from discreteness  of space
\cite{bun05}.

In  conclusion, we  believe  our  work opens  a  possible approach  to
realize  Wheeler's  phrase  ``it  from bit'',  namely,  that  physical
reality  derives its  existence  from a  deeper information  theoretic
layer   \cite{hweo},   which   we   have   called   the   sub-physical
universe/matrix.   To quote  Wheeler, ```It  from bit'  symbolises the
idea that every item  of the physical world has at bottom  - at a very
deep bottom, in most instances - an immaterial source and explanation;
that which we call reality arises in the last analysis from the posing
of yes-no questions and the registering of equipment-evoked responses;
in short,  that things physical are  information-theoretic in origin.'
Perhaps the most  surprising aspect of our proposed  model is that the
possible  profound connection  between physics  on the  one  hand, and
information theory and computer science on the other, that it suggests
finds a  very commonplace manifestation--  in the classicality  of the
familiar macroscopic world.

I am thankful to Prof. B. Iyer, Drs. Madhavan Varadarajan and S. Surya
for useful comments and discussions.


\begin{thebibliography}{100}
\bibitem{gring} A Gaussian integer is a complex number of the form
$a + b\iota$, where $a, b$ are integers. The set $S$ of
all Gaussian integers is a subring of $\mathbb{C}$.
\bibitem{segfu} Segmentation fault is an error that 
occurs when a computer program attempts to access memory
outside the area allocated to it.
\end{thebibliography}
\end{document}